\documentclass[epjST,final]{svjour}
\usepackage{graphics}
\usepackage{url}
\begin{document}
\title{Introduction to colloidal dispersions in external fields}
\author{H. L\"owen}
\institute{Institut f\"ur Theoretische Physik II: Weiche Materie, 
Heinrich-Heine-Universit\"at D\"usseldorf, 
40225 D\"usseldorf, Germany}
\abstract{ %abstract
Progress in the research area of colloidal dispersions in external 
fields within the last years is reviewed.
Colloidal dispersions play a pivotal role as model systems for phase transitions in classical 
statistical mechanics. In recent years the leading role of colloids to realize model 
systems has become evident not only
for equilibrium situations but also far away from equilibrium. 
By using external fields (such as shear flow,
electric, magnetic or laseroptical fields as well as confinement), a colloidal suspension can be brought
into nonequilibrium in a controlled way. Various kinds of equilibrium and nonequilibrium phenomena
explored by colloidal dispersions are described providing also a guide and summary to this special 
issue. Particular emphasis is put on the comparison of real-space experiments, computer simulations and
statistical theories.
} %end of abstract
\maketitle
\section{Introduction}
\label{Introduction}

The special topic of this themed issue is
the physics of {\it colloidal dispersions} which is a rapidly growing research area in the
interdisciplinary soft matter domain. Colloidal dispersions are solutions of mesoscopic
solid particles with typical sizes ranging from 1--10 $nm$ to about 1--10 $\mu m$
 with a stable (i.e.\ non-fluctuating) core embedded in a molecular fluid solvent.
Among the various soft matter systems, colloidal dispersions play a pivotal role
as model systems,  as they can be both prepared and characterized in a controlled way. The effective
interaction between the colloidal particles can be tailored by changing, e.g., the salt concentration in
the solvent. Moreover, colloidal suspensions can be regarded as the simplest prototype of soft matter:
the length scale separation between the molecular solvent and the mesoscopic particles is unique and
complete. Spherical particles without any additional structure on the mesoscopic length scale possess
the simplest and highest possible symmetry. This directly implies that a simple theoretical modelling
of a single particle without many fitting parameters is possible. Exciting questions concern collective
many-body effects induced by cooperation and self-organization of many particles,
in particular for bulk phase transitions such as fluid-fluid phase
separation, freezing as well as glass and gel formation.

\begin{figure}
  %\begin{minipage}{\linewidth} 
    %\renewcommand{\footnoterule}{}
    \centering
    \resizebox{0.75\columnwidth}{!}{\includegraphics{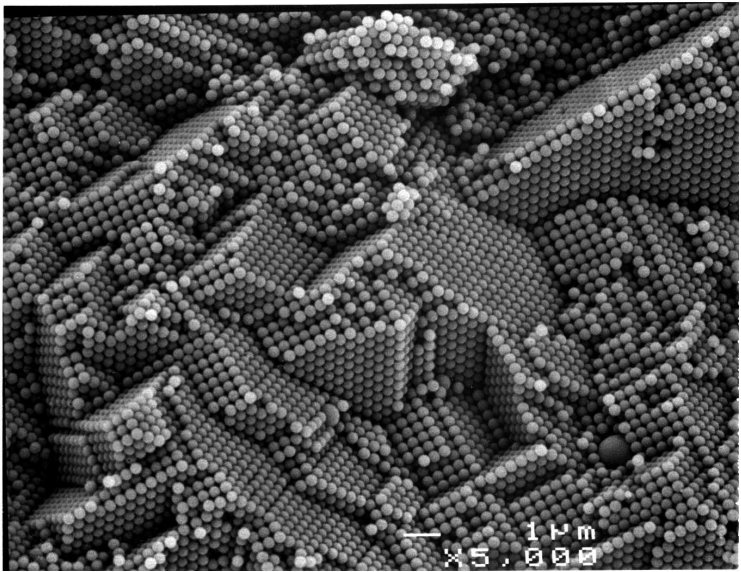}}
    \caption{Real-space image of submicron colloidal PMMA spheres will exhibit long-range positional order on a crystalline lattice.\protect\footnotemark }
  \label{fig:1}
\end{figure}

As an example, Fig.~\ref{fig:1} shows an electron micrograph of mesoscopic colloidal particles
which look really like spheres on the micron scale. What can be directly seen from Fig.~\ref{fig:1}
is that the particles form crystalline lattices with a mesoscopic lattice constant.
Correspondingly, the time scale upon which crystallization occurs 
is slow enough such that configurational changes can be studied by tracking the individal particle trajectories.
Therefore the full "microscopic", i.e.\ particle-resolved, information is available, 
similar as in classical particle--resolved computer simulations.

\footnotetext{Taken from \protect\url{http://www.physics.nyu.edu/pine/Pine___Res___Clusters.html}}    

Moreover, the properties of colloidal dispersions can efficiently be tailored and controlled by
employing {\it external fields}. The study of of an external control via external
fields has two main reasons: 

\begin{enumerate}
  \item First, by definition, soft matter reacts sensitively ("softly") upon external 
perturbations and manipulations. The occurrence of stable colloidal bulk 
samples is the exception rather than the
rule, i.e., one has to protect the sample carefully against shear and other perturbations. How protection can work needs therefore detailed study.

  \item The second reason is that strong external fields induce qualitatively novel effects not known from equilibrium bulk. Complementarily, the softness is rather exploited than being protected in this case.
\end{enumerate}

An external field can either give rise to an equilibrium
problem (e.g. by modifying the interparticle interactions) or drives the system into
nonequilibrium. Colloidal dispersions in external fields play therefore a similar pivotal role
as model systems for  controlled nonequilibrium situations. This
opens the way for a fundamental understanding of nonequilibrium phenomena
which is important also for technological applications. Significant achievements
in understanding the principles of the behaviour in non-equilibrium were gained
by a systematic comparison of results from experiment, theory and computer
simulation. In this special issue, the structural
and dynamical behaviour of colloidal dispersions in various external fields
are discussed  both in equilibrium
and in non-equilibrium. The external fields can be:

\begin{itemize}
\item shear flow

\item electric fields

\item laser-optical and magnetic fields

\item confining geometries
\end{itemize}

During the last years, significant developments have stimulated
the colloid research in general and colloidal physics in external fields in particular.
First of all, qualitative novel behaviour genuinely induced by an applied field has been
discovered in theory and experiment leading to new nonequilibrium phenomena. Second,
different experimental techniques have been applied to study colloidal dispersions,
in particular real space techniques.
Next, computer simulation schemes have been applied and developed, e.g.\ in order to simulate
hydrodynamic interactions between the colloidal particles mediated by the
solvent. Last but not least, new statistical theories for colloidal models have been 
developed and significantly extended.

\begin{figure}
  \centering
  \resizebox{0.9\columnwidth}{!}{\includegraphics{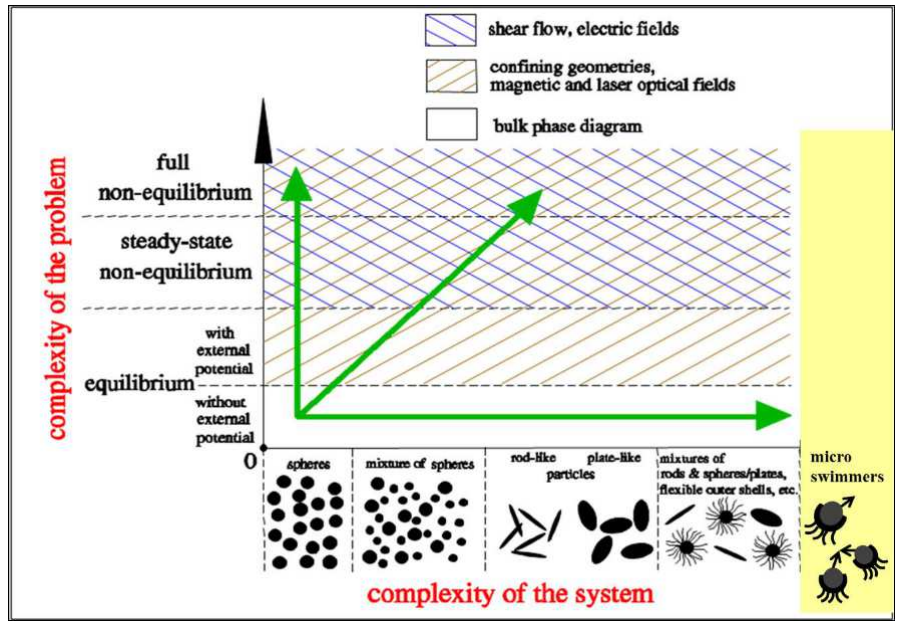}}
  \caption{Road map of complexity for colloidal dispersions in external
fields: while the x-axis shows the complexity of the system, the y-axis shows the
complexity of the problem. Regions which can be accessed by different kinds of external
fields are indicated. The arrows indicate recent research directions. Active particles or microswimmers, which exhibit a special complexity induced by their internal propulsion, are also indicated.}
  \label{fig:2}   
\end{figure}

The major motivation is a “microscopic” understanding - on the basic time- and length
scales - of the particles in non-equilibrium. The route toward this challenging goal
can be best explained in terms of complexity. The level of different complexities
is classified and summarized in the traditional complexity diagram of Fig.~\ref{fig:2}
\cite{Lowen2001}. On the horizontal
axis, the degree of complexity of the system is shown related to the statistical
degrees of freedom present in the system. The simplest case are spherical particles
while mixtures and orientational degrees of freedom represent a higher level of
system complexity. Even higher complexity is arising from particles with a nonconvex
shape like colloidal molecules or with internal degrees of freedom, e.g.
with an internal motor \cite{Romanczuk2012,Cates2012}. 
On the other hand, the complexity of the question asked
which is shown on the vertical axis can comprise equilibrium situations, such
as inhomogeneous systems near walls and in restricted geometries, steady-state
non-equilibrium cases (such as systems under permanent shear flow or other time-
independent or oscillatory external fields). Finally, full non-equilibrium situations
arise if a field is turned on or switched off.
A thorough understanding can only be achieved via a systematic, step-by-step
process in the complexity diagram as indicated by the arrows in Figure 2.
While the covering of the unknown part was a vision more than 10 years ago 
when the complexity diagram was first proposed \cite{Lowen2001}, 
this vision has now become true and was realized as documented by the research 
results contained in this special issue. 

In this introductory article, recent highlights obtained in our understanding of colloidal dispersions
in various external fields will be briefly presented. First we shall consider four different types 
of external fields consecutively, namely  shear flow, electric, laser-optical and magnetic 
fields as well as confinement. Then we shall give an outlook about future research activities 
in the realm of colloid physics where we shall address anisotropic colloids and self-propelled particles with 
intrinsic degrees of freedom, in particular. This article therefore provides also a brief classification
of the 23 following minireviews contained in this special issue.

A large portion  of the results discussed here were obtained 
within the  German-Dutch Collaborative Research Centre
SFB TR6 with the title {\it Physics of Colloidal Dispersions in External Fields} within 
the last funding period 2009-2013. The SFB TR6 network  comprises the locations 
D\"usseldorf, J\"ulich, Konstanz, Mainz and Utrecht but has various collaborations with 
more external project partners.

As a general final remark, colloidal dispersions show strong analogies to complex plasmas
which are similar form-stable mesoscopic particles embedded in a plasma. The similarities 
and complementarities were recently discussed, see e.g.\ \cite{ivlev2012complex} for an overview.
Examples include  separation kinetics in binary mixtures \cite{Wysocki10}
and lane formation \cite{Sutterlin09,vissers2011} which are found both in colloids
 and in complex plasmas but with  different details in the dynamics. Another fruitful link is to granular materials \cite{Aranson} 
which allow for a direct observation of the particles but do not 
exhibit Brownian motion.

\section{Shear fields}
\label{ShearFields}
A shear field naturally occurs for colloidal dispersions in a solvent flow.
In the past years, there has been a significant progress in understanding sheared 
colloidal suspensions within a microscopic or schematic mode-coupling-like theory,
in particular close to the glass transition. For example, a
schematic model for the constitutive rheology of glasses was presented \cite{Brader09}. 
The resulting tensorial structure of the schematic model satisfies applicable
 invariance laws for the nonlinear flow of materials without inertia, 
 such as colloidal suspensions. As a concrete example, 
 the full dynamic yield stress manifold was calculated by addressing a family of steady 
 flows that interpolate between planar and uniaxial elongation. The resulting
 yield surface is quite similar, but not identical, to the empirical form of 
 von Mises that has been widely used to model static yield and plasticity in 
 solids. This theory thus gives a first--principle justification for this 
 empirical law. 
A variant of this theory was tested against nonlinear rheology data of concentrated 
thermosensitive core-shell particles and excellent  agreement was found 
concerning yielding, shear-thinning and large amplitude oscillatory shearing 
\cite{Siebenbuerger}. More aspects are reviewed in \cite{EPJEST_6}.

 Soft particles
deform under shear and their structural and dynamical behaviour was intensely studied both by
experiment \cite{EPJEST_3} and theory \cite{EPJEST_4}. For instance, the dynamical 
deformation of ultrasoft colloids as well as their dynamic frictional forces
 were numerically investigated, when one colloid is dragged past another at constant velocity \cite{Singh}.
 Hydrodynamic interactions are captured by a particle-based mesoscopic simulation method. 
 At large drag velocities, an apparent attractive force was found for departing colloids along
 the dragging direction. 

Transient dynamics after switching on shear was obtained
by a joint venture of theory, simulation and experiment, see e.g.\ \cite{Laurati_CodefJPCM}. In particular, 
concentrated hard-sphere suspensions and glasses were investigated with rheometry, confocal microscopy, 
and Brownian dynamics simulations during start-up shear, providing a link between microstructure, 
dynamics, and rheology \cite{Koumakis}. The microstructural anisotropy is manifested in the extension axis where 
the maximum of the pair-distribution function exhibits a minimum at the stress overshoot. The interplay
 between Brownian relaxation and shear advection as well as the available free volume determine the 
structural anisotropy and the magnitude of the stress overshoot. 

Colloids provide
the fascinating possibility to drag single particles through the suspension, which gives
access to microrheology (as opposed to macrorheology where macroscopic boundaries
are moved). Several theoretical aspects of microrheology were discussed 
\cite{Harrer_CodefJPCM,Ryan_CodefJPCM_part1,Ryan_CodefJPCM_part2} including again a schematic 
model of mode-coupling theory \cite{Gnann}.  The latter
model describes the strongly nonlinear behavior of the microscopic
 friction coefficient as a function of applied external force in terms of a delocalization transition, 
 which is visible in Brownian dynamics simulations of a system of quasi-hard spheres and experimental
 data on hard-sphere-like colloidal suspensions. Moreover,  molecular dynamics computer simulation of a glass-forming Yukawa mixture was used 
to study the anisotropic dynamics of a single particle pulled by a constant force \cite{Winter}. 
Beyond linear response, a scaling regime was found where a force-temperature superposition
 principle holds. This is summarized in the minireviews  \cite{EPJEST_5,EPJEST_7}.

Finally, total Internal Reflection Fluorescence Cross-Correlation Spectroscopy 
uses fluorescent colloidal particles to probe 
the properties of a flow field near a surface \cite{Schmitz2011}. 
Careful experiments combined with a detailed 
theoretical analysis employing large-scale computer simulations are able to 
measure properties like the slip length accurately. However, the results also clearly show that the method requires detailed modeling 
 of the diffraction phenomena at the confocal microscope's objective, and that 
this aspect needs further improvement.

Finally anisotropic particles show an even more complex response to shear flow as isotropic ones.
Some recent findings are summarized in \cite{EPJEST_2}.

\section{Electric fields}
\label{ElectricFields}

If a charged colloidal particle is placed into an electric field it gets into motion. 
This electrokinetic effect is, however, nontrivial due to the presence of counterions 
in the solution and due to solvent flow effects. In fact, electrophoresis constitutes 
already a nontrivial problem for a single colloidal particles
since electrostatics and hydrodynamics couple in a complex way. Accordingly, a 
collective ensemble of charged colloids responds in an  even more complicated 
way to an external electric field.
In this special issue, electrokinetics will be accessed by experimental studies \cite{EPJEST_8,EPJEST_11} 
by computer simulations \cite{EPJEST_10,EPJEST_13} and by statistical theory \cite{EPJEST_9}.

An electric field can also be used to manipulate and steer the self-assembly of colloids. 
This idea is in particular useful for nonspherical particles. Applications of these methods 
are discussed in \cite{EPJEST_12}.

Finally we mention that binary mixtures of oppositely charged colloids can also be 
prepared corresponding to a mesoscopic realization of a molecular molten 
salt \cite{leunissen2005}. The advantage of the colloidal system is that arbitrary charge and size ratios can be realized. This mixture 
exhibits fascinating bulk phase diagrams with a wealth of different mixed crystals \cite{ivlev2012complex}.
When placed into a DC electric field the oppositely charged particles will move against each other.
At high enough field strength, lane formation of like-charge particles occurs
\cite{Dzubiella02,leunissen2005,lowen2010instabilities}.

More recently, real-space experiments, computer simulation and analytic theory were performed for lane formation of oppositely charged colloids in electric fields \cite{vissers2011,KohlJPCM2012,GlanzJPCM2012}. 
The results demonstrate a continuous
 increase in the fraction of particles in a lane in the case that oppositely charged particles 
are driven by 
 an electric field. This behavior is accurately captured by Brownian dynamics simulations. 
By studying the 
 fluctuations parallel and perpendicular to the field a mechanism was identified that 
underlies the formation
 of lanes. In an AC-electric field, bands perpendicular to the field directions form.
The detailed nonequilibrium phase diagram depends on the amplitude and frequency of 
the applied field as predicted by computer simulations \cite{Wysocki_PRE_2009}.
The banding effect was confirmed by real-space experiments \cite{Vissers}. However, a detailed understanding of the hydrodynamic interactions between colloids \cite{Wysocki_JPCM_2011,lowen2010instabilities} is still lying ahead.

\section{Laser-optical and magnetic fields}
\label{LaserOpticalMagneticFields}

Laser-optical fields can be used to tailor a random substrate potential for colloids
\cite{Mazilu_CodefJPCM} or to bind colloids optically \cite{Dillmann_CodefJPCM}. A modulated substrate potential 
can be created by superimposing various laser fields and then the phase 
transitions of colloids in these periodic 
substrates can be explored where, in general, a large deviation from the bulk behaviour is found.
The minireview \cite{EPJEST_19} of this special issue summarizes some important examples.

External magnetic fields are typically used to
create dipolar repulsions of colloids pending at an air-water interface. This provides
an avenue to two-dimensional systems, where the freezing transition \cite{Wilms_CodefJPCM} and various
transport phenomena through channels are in the focus of recent research \cite{Kreuter_CodefJPCM,Malijevsky_CodefJPCM}.
One example is the realization of a sudden quench which can be
applied to binary mixtures of superparamagnetic 
colloidal particles confined at a two-dimensional water-air interface 
by a sudden increase of an external magnetic field \cite{assoud2009}. This quench realizes
 a virtually instantaneous cooling which is impossible in molecular systems.
Another example are two dimensional fluids with quenched disorder as realized by particles pinned to the substrate which show significant deviation in phase behavior as compared to pure systems \cite{Deutschlaender13}.
 
Various aspects of the  impact of magnetic fields on colloids is intensely discussed 
in five minireviews of this special issue: The motion of magnetic particles 
in external potential barriers
\cite{EPJEST_14} and in time-dependent magnetic fields \cite{EPJEST_15}, 
the superposition of a magnetic field with shear flow \cite{EPJEST_18} 
and gravity \cite{EPJEST_16}
and the extension from one-component systems to binary mixtures \cite{EPJEST_16}.

\section{Confinement}
\label{Confinement}

The classic set-up of confinement is a slit-geometry realized by 
squeezing the suspensions between two parallel glass plates. The plate separation can easily be
an interparticle spacing such that the effects of confinement are drastic as compared to the bulk.
An important question concerns the nature of bulk phase transition like freezing, 
condensation or the glass transition in confinement.

Colloidal crystal confined in slit geometry have structures widely different from their bulk. 
A survey of recent developments can be found in the minireview \cite{EPJEST_20}.
Examples include multilayered structures \cite{leunissen2005} and new structures solely induced by confinement 
 \cite{Oguz_2012_PRL,Moire}.  Also the dynamics of crystallization in confinement has been studied
in detail, for example under gravity  \cite{Moire} and in a flow-field \cite{Reinmueller2012}.

The dynamics of sedimentation of an initially inhomogeneous distribution of hard-sphere colloids 
confined in a slit was studied by simulation and real-space microscopy experiments \cite{WysockiSM2010,wysocki2009}.
The scenario of the observed Rayleigh-Taylor-like instability depends crucially on the 
hydrodynamic interactions mediated by the solvent.

The glass transition in slit geometry was calculated by
extending mode-coupling theory, a microscopic theory for the glass transition of liquids \cite{Lang}. The glass transition line was calculated as a function of the distance of the 
 plates for the case of a hard sphere fluid and an oscillatory behavior was obtained as a result of the structural changes related to layering.

In cylindrical confinement,
computer simulations of colloid-polymer mixtures \cite{Wilms} 
have shown the existence of two distinct rounded transitions for condensation. 
The transition from multi-domain to a single domain
 state explains the hitherto mysterious "hysteresis critical point".
Moreover, helical crystalline structures were predicted in cylindrical confinement \cite{Oguz_2011}.

Colloids can also be confined to liquid-gas interface. Capillary 
waves then lead to a mutual lateral attraction between the colloidal 
particles. The actual form of the attractive force is
formally analogous to two-dimensional  screened Newtonian gravity with the capillary length
as the screening length. Therefore the set-up has the same equations 
as cosmology albeit on a much smaller length scale. The "gravitational" collapse was recently explored 
using Brownian dynamics simulations, density functional theory, and analytical perturbation theory
\cite{Bleibel}, see also the minireview \cite{EPJEST_24}.

Finally, anisotropic particles in confinement possess even more possibilities of
structural phase transitions due to a competition of many length scales. 
Some aspects are reviewed in the subsequent papers \cite{EPJEST_21,EPJEST_22,EPJEST_23}.

\section{Non-spherical particle shapes}
\label{NonSpherical}

By now a plethora of non-spherical particle shapes can be prepared, 
see \cite{ivlev2012complex} for a recent review. Frequently these particles are 
called "colloidal molecules", in analogy to the traditional molecules in the 
microscopic regime. The advantage of the colloids relative to real molecules, 
however, is that arbitrary shapes can be principle be prepared at wish.

Here we mention three key examples: first, the phase behaviour of
boardlike colloids  as realized
 by goethite (a-FeOOH) particles were explored and biaxial nematic and 
biaxial smectic phases were found \cite{van2009experimental}. The macroscopic domains were oriented 
by a magnetic field and their structure was revealed by small angle x-ray 
scattering. Second, particles with a different chirality 
(enantiomers) were considered \cite{Meinhardt}.
For many applications it is necessary to separate particles with different chirality.
Such a separation can be achieved in microfluidic or nanofluidic channels if a solvent
flow field is used  which breaks chiral symmetry and has regions with high local shear \cite{Meinhardt}.
Such flow profiles can be generated in channels confined by walls with different hydrodynamic 
boundary conditions (e.g., slip lengths). Due to a nonlinear hydrodynamic effect, particles 
with different chirality migrate at different speed and can be separated. 

Third, the theory of Brownian dynamics though well-developed for spheres and rods
was completely formulated for interacting rigid biaxial particles with an arbitrary shape
\cite{Wittkowski2012,Wittkowski2012MolPhys}.
Regarding more recent advances for anisotropic particles we refer the reader to 
the papers \cite{EPJEST_21,EPJEST_23} of this special issue.

\section{Active Colloids}
\label{ActiveColloids}

In contrast to passive particle in an external field,
self-propelled particles dissipate energy and move autonomously. Their propagation 
direction is thereby an inner degree of freedom which is not only subjected to thermal fluctuations, the external field and particle interaction, as opposed to passive particles.
This leads to intrinsic nonequilibrium behaviour
\cite{Romanczuk2012,Cates2012}.  Self-propelled colloidal particles have been 
prepared where the external field again plays a leading role for the self-propagation. 
As prominent examples we mention catalytically-driven Janus-particles \cite{erbe2008various}
and thermally driven colloids in a phase-separating solvent \cite{Bechinger,KuemmelPRL2013,Buttinoni2013_PRL}.

Self-propelled rod-like particles exhibit clustering which is not present in equilibrium 
\cite{Wensink_Lowen_2008,Marceau_Gompper}. They can be efficiently caught in a wedge-like trap
\cite{Kaiser} and show turbulence-like behaviour at intermediate densities \cite{Wensink_PNAS}.
This turbulent behaviour occurs at low-Reynolds number in contrast to 
the traditional turbulence which happens at high Reynolds number.

Also the freezing transition has been studied for self-propelled colloids \cite{Bialke}.
The transition differs significantly from bulk freezing.
 The transition is accompanied by pronounced structural
 heterogeneities. This leads to a transition region between liquid and solid in which
 the suspension is globally ordered but unordered liquidlike "bubbles" still persist.
For large propagation speed there is a transition from a resting crystals to a travelling crystal
which migrates collectively \cite{Menzel}. In terms of theory, the microscopic density functional 
approach has been formulated also for active systems \cite{Wittkowski2012MolPhys}.
Finally, the articles \cite{EPJEST_21,EPJEST_14} of this special issue discuss 
more recent findings for self-propelled colloidal particles.

\section{Conclusions}
\label{Conclusions}

Meanwhile colloidal suspensions are not only excellent model systems to
understand  bulk phase transitions on the particle scale. The past decade 
has shown that colloids also provide principle guidance to understand and classify
nonequilibrium behaviour as induced by external fields. The present special issue 
discusses a plethora of examples for field-induced effects in colloidal dispersions.
New research typically goes hand-in-hand by using complementary methods, namely
real-space experiments, computer simulations and statistical theories.

The possibility of synthesize novel particle shapes and to prepare self-propelled
colloidal particles has given two recent major boost for the research field of colloids.
The collective behaviour of-non-spherical, self-propelled and active particles \cite{Haenggi2009_Rev} 
is expected to become a flourishing field for the future.

\begin{acknowledgement}
  Discussion with all members of the SFB TR6 are acknowledged.
  This work was supported by the DFG within SFB TR6.
\end{acknowledgement}

%%%References
\bibliography{codef_refs}

\end{document}